%% file: main.tex
\documentclass[a4paper,11pt]{article}
\usepackage[T1]{fontenc} 
\usepackage[utf8]{inputenc}
\usepackage{microtype}
\usepackage{lmodern}
\usepackage[english]{babel}
\usepackage{siunitx}
\usepackage{xspace}
\usepackage{pos}

\usepackage[export]{adjustbox}  

\title{Air-Shower Radio Simulations -- Where we stand and where we go}
\ShortTitle{Air-Shower Radio Simulations}

\author*[a,b]{Tim Huege}

\affiliation[a]{Karlsruher Institut für Technologie, Institut für Astroteilchenphysik (IAP), Karlsruhe, Germany}
\affiliation[c]{Vrije Universiteit Brussel, Astrophysical Institute, Brussels, Belgium}
\emailAdd{tim.huege@kit.edu}

\abstract{
Simulations of the radio emission from extensive air showers have been key in establishing radio detection as a mature and competitive technique. In particular, microscopic Monte Carlo simulations have proven to very accurately describe the emission physics and are at the heart of practically all analysis approaches. Yet with new applications -- for example very inclined air showers, cross-media showers, extreme antenna densities, and higher-frequency measurements -- come new challenges for accurate and efficient simulations. I will review the state of the art of the existing simulation approaches and discuss where further improvements might be needed and how they can be achieved.
}

\FullConference{%
  Acoustic \& Radio EeV Neutrino Detection Activities (ARENA)\\
  7–10 Jun 2022\\
  Santiago de Compostela, Spain
}

\begin{document}
\maketitle

\input{definitions}


\section{Introduction}

Simulations of the radio emission from extensive air showers have been a decisive factor in establishing the radio detection technique for extensive air showers \cite{HuegePLREP}. The approach relied most upon is the microscopic simulation of the radio emission from single electrons and positrons in the air-shower cascade using the ``endpoints formalism'' \cite{JamesFalckeHuege2012} in the CoREAS \cite{Huege:2013vt} code and the time-domain version of the ZHS formalism \cite{ZasHalzenStanev1992} in the ZHAireS \cite{AlvarezMunizCarvalhoZas2012} code.\footnote{The endpoints formalism describes the radio emission as arising from the acceleration in the ``kinks'' between the straight track segments used to approximate the continuous motion of a charged particle, whereas the ZHS formalism associates the emission with the straight track segments themselves.} The predictions by these simulations have been tested with data from many different experiments, so far without revealing any deficiencies.

In this article, I will shortly review the degree of certainty we have achieved on these simulation predictions, first for the case of vertical air showers with zenith angles up to 60$^\circ$, then for the case of inclined showers with zenith angles above 60$^\circ$. Furthermore, I will discuss future directions in which more work should be invested to cope with the ever-increasing need for more complex and accurate simulation scenarios.


\section{Simulations for vertical air showers}

Radio emission from vertical air showers, i.e., those with zenith angles up to $\sim 60^\circ$, has been measured with many different experiments, and their data have been compared extensively with microscopic simulations.

\begin{figure}
\centering
\includegraphics[width=0.49\textwidth]{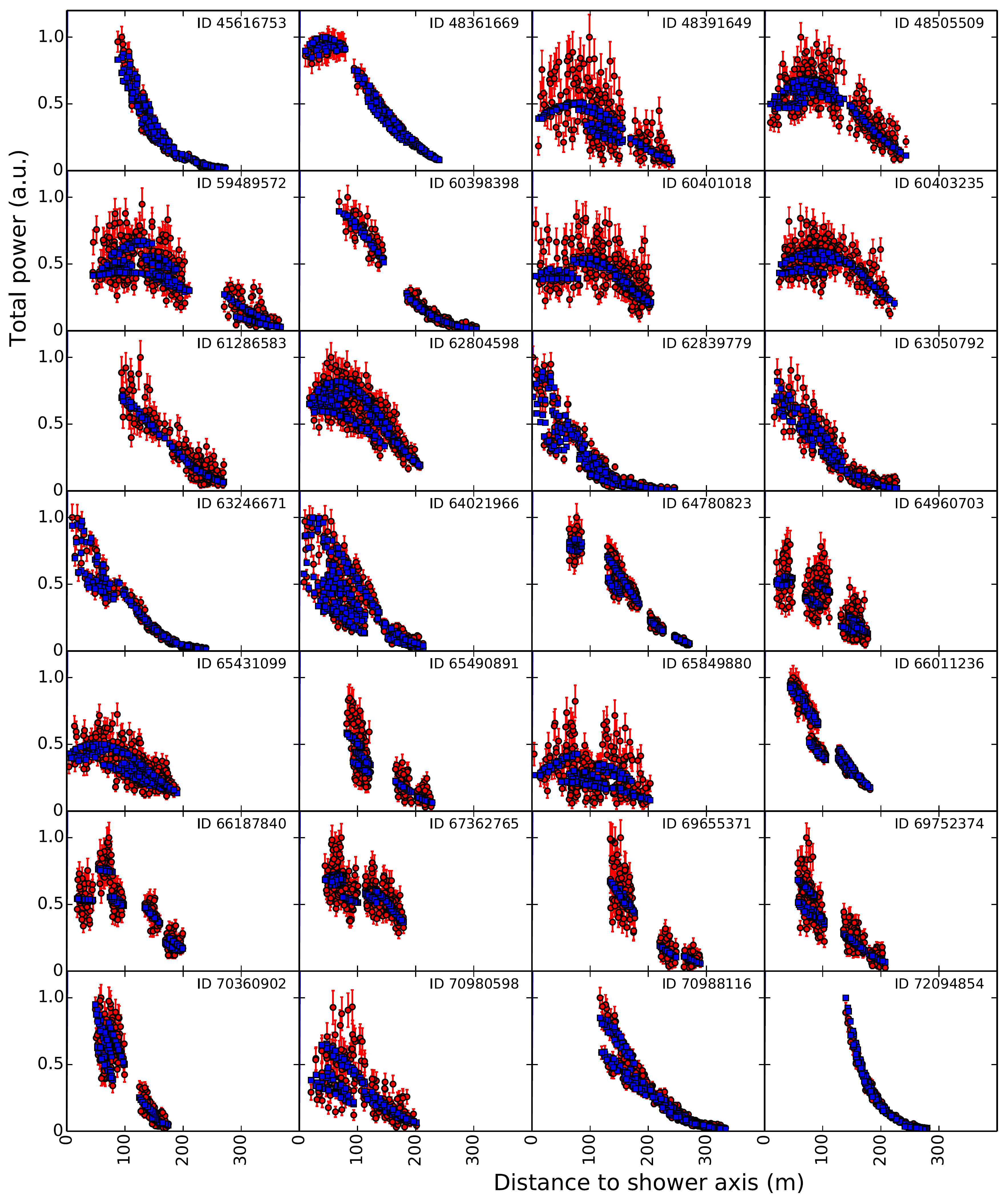}
\includegraphics[width=0.49\textwidth]{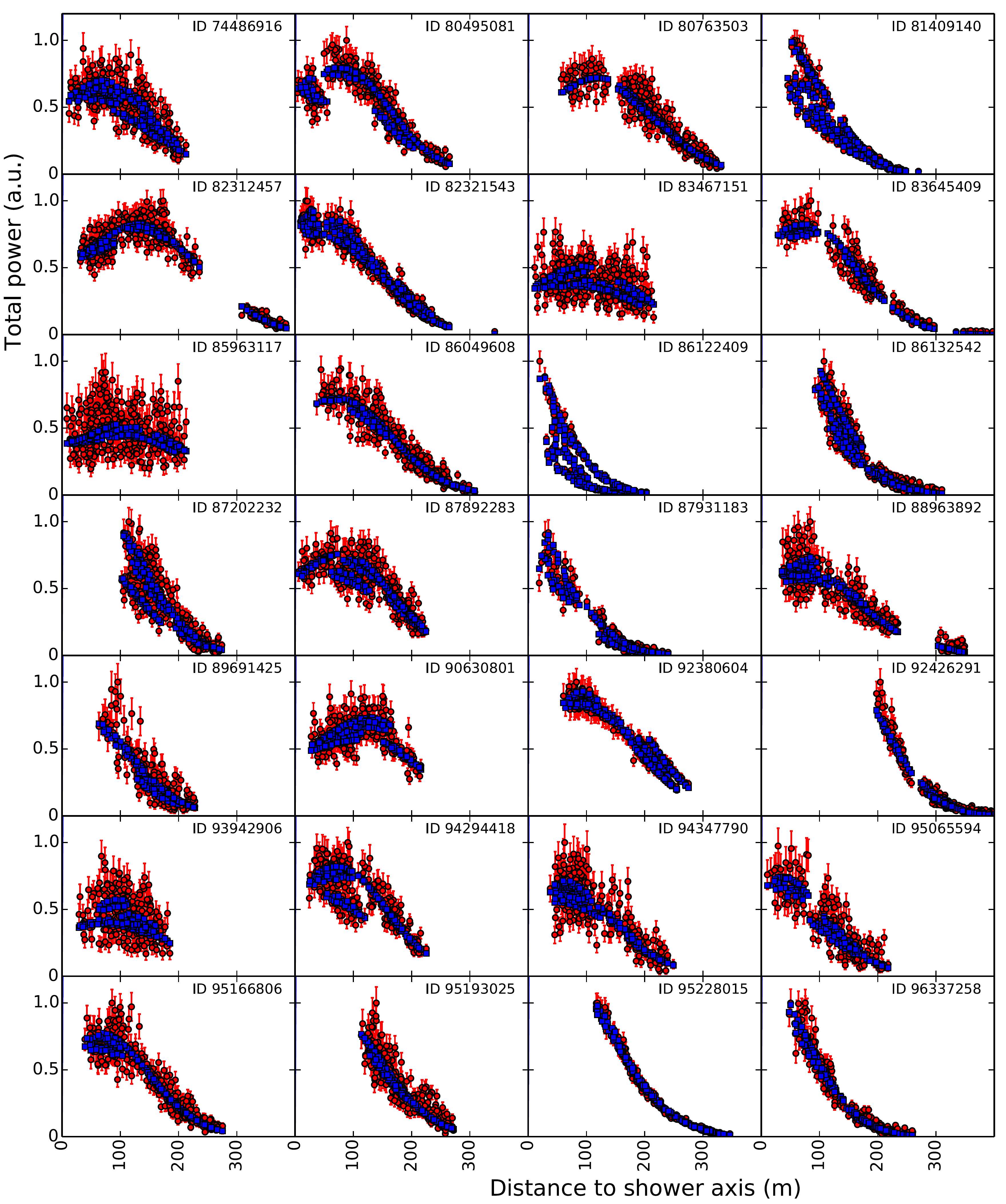}
\caption{\label{fig:lofarldfs}Lateral distributions of the time-integrated power measured by LOFAR for a subset of 48 of their measured events (red circles) as compared with the best-fitting CoREAS simulations (blue squares). From \cite{LOFARNatureXmax}.}
\end{figure}

\begin{figure}
\centering
\includegraphics[width=0.9\textwidth]{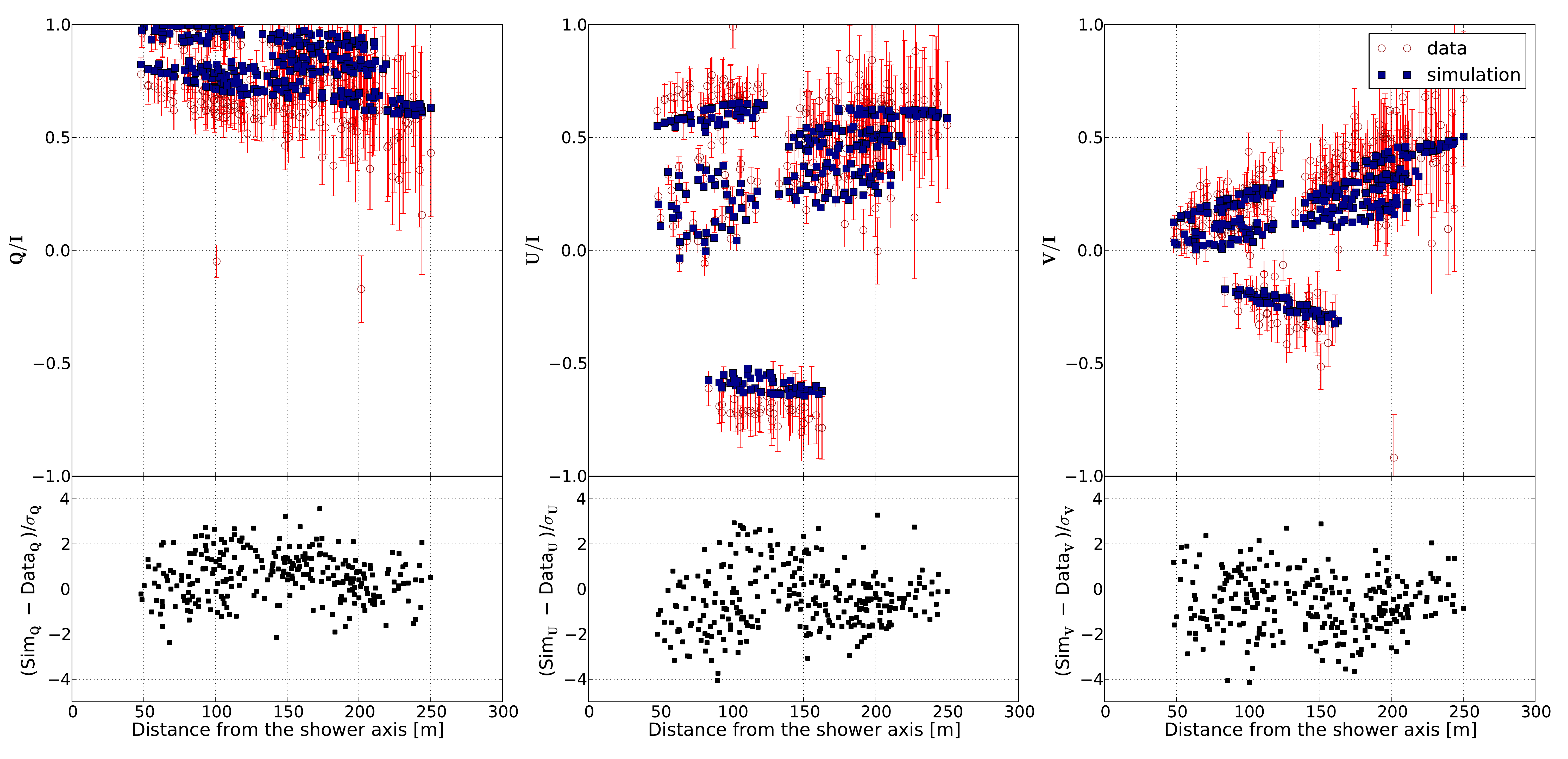}
\caption{\label{fig:lofarcircularpol} Comparison of measured (red circles) and CoREAS-simulated (squares) Stokes parameters for one LOFAR event. $Q/I$ (left) and $U/I$ (middle) quantify the degree and orientation of the linear polarization of the signal, $V/I$ (right) denotes the degree of circular polarization.  From \cite{Scholten:2016gmj}.}
\end{figure}

The most in-depth comparison of simulations and data has been performed by LOFAR, which has measured several hundred air showers with several hundred antennas each \cite{LOFARNatureXmax}. While the nominal frequency band quoted for the LOFAR low-band antennas is 30--80~MHz, one has to keep in mind that the antennas are highly resonant dipoles and thus effectively only measure in a small band with a few MHz width near 60~MHz. Nevertheless, the agreement between CoREAS simulations and data is impressive, as illustrated by the lateral signal distributions in terms of time-integrated power (the equivalent to energy fluence on the level of voltage traces, i.e., with the detector response still present) shown in Fig.\ \ref{fig:lofarldfs}. These comparisons incorporate a per-event characterization of the atmospheric density and refractive index profiles (including humidity effects) at the time of measurement, as derived from GDAS \cite{Mitra:2020mza}. In addition to  the total signal strength, also the polarization characteristics match precisely between CoREAS simulations and LOFAR data, even down to the degree of circular polarization, as shown in Fig.\ \ref{fig:lofarcircularpol}.

Experiments like Tunka-Rex \cite{Tunka-Rex:2015zsa} and the Auger Engineering Radio Array (AERA) \cite{AERAEnergyPRL}, probing the full 30--80~MHz band with an approximately uniform response, have also made comparisons of CoREAS simulations and data. In both cases, simulations and data agreed well within systematic uncertainties \cite{HuegePLREP}. With AERA data, it was confirmed that the absolute strength of the emission predicted by CoREAS as well as ZHAireS is in agreement with the air shower parameters determined by the surface detector \cite{AERAEnergyPRL}, the energy scale of which has been set with fluorescence detection. Unfortunately, the systematic uncertainty of this absolute scale comparison is currently relatively large, but bound to improve in the future. With AERA data, it has also recently been shown that the determination of the depth of shower maximum, $X_\mathrm{max}$, using comparisons of CoREAS simulations and data, yields results comparable with measurements with the Auger Fluorescence Detector \cite{PontXmaxARENA2022}, another confirmation that the simulations are reliable.

\begin{figure}
\centering
\includegraphics[width=0.49\textwidth]{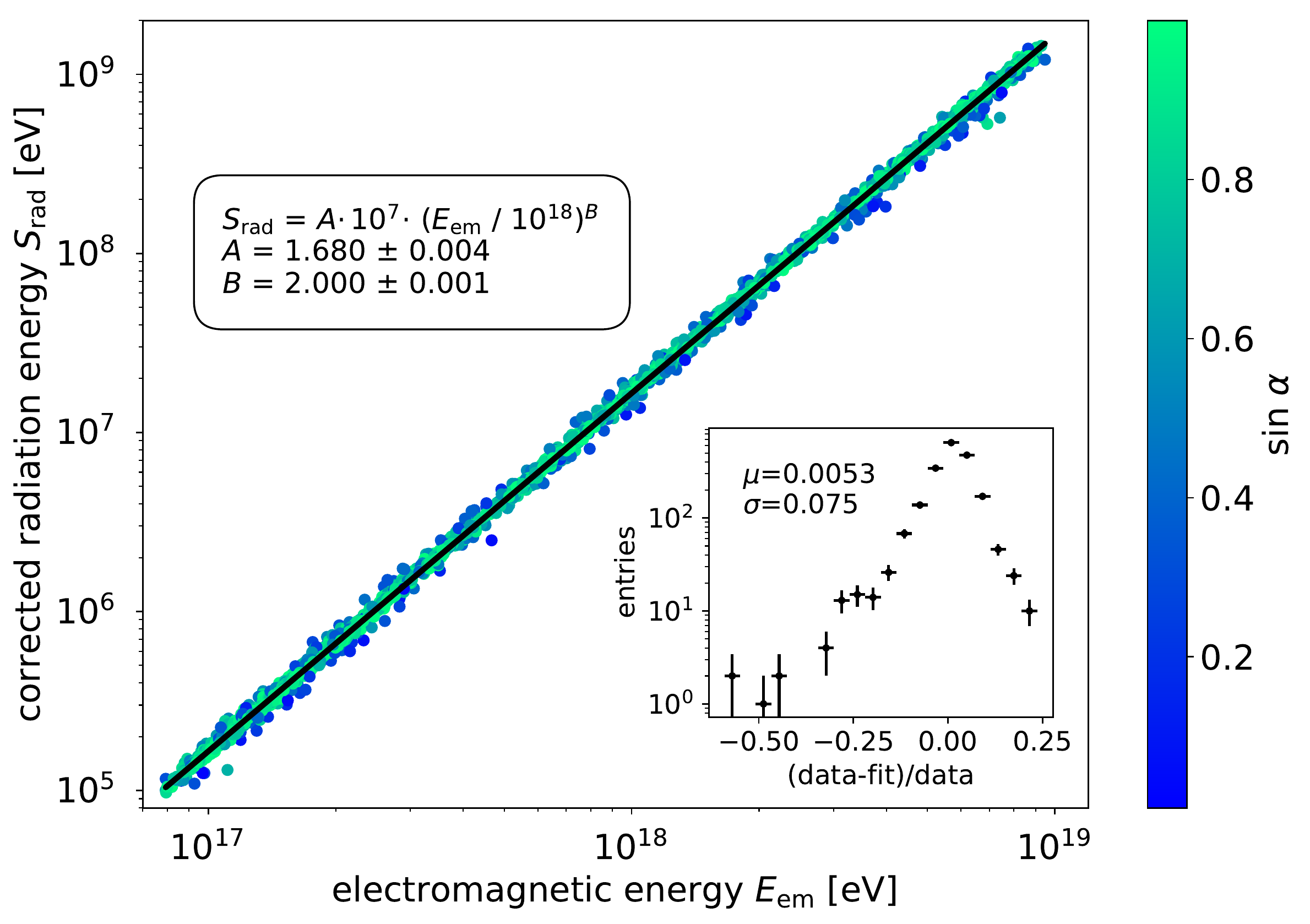}
\includegraphics[width=0.49\textwidth]{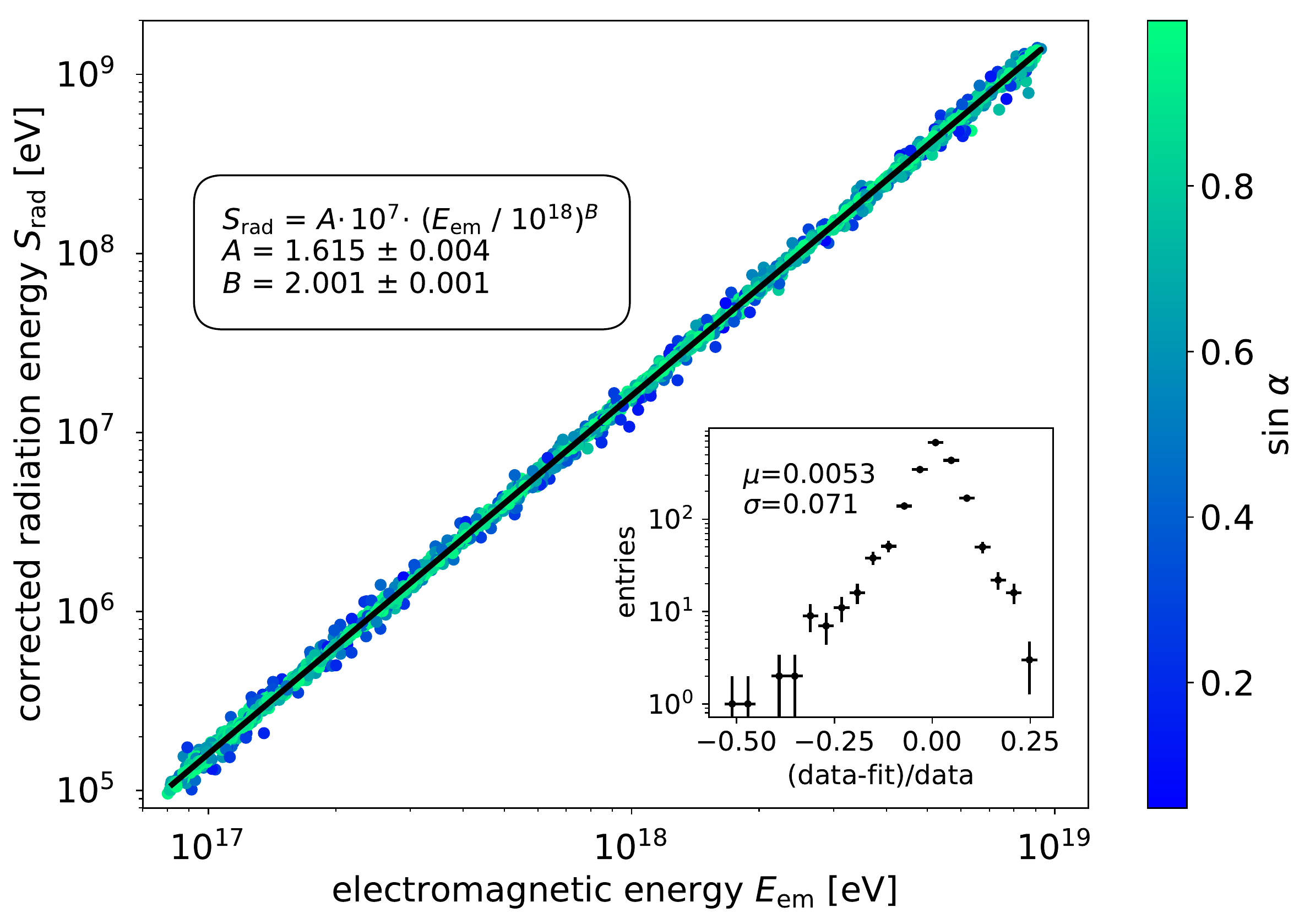}
\caption{\label{fig:coreaszhaires} Correlation of the corrected radiation energy with the electromagnetic energy in the 30--80~MHz band for CoREAS (left) and ZHAireS (right). (Corrected radiation energy corresponds to the total energy deposit on the ground in the form of radio waves corrected for air-density effects and magnetic field orientation.) The absolute scales of the simulations agree well. From \cite{Gottowik:2017wio}.}
\end{figure}

Simulations with ZHAireS have been shown to agree very well with CoREAS simulations in terms of the total emitted radiation energy, see Fig.\ \ref{fig:coreaszhaires}, with a shift of only 5.2\% in corrected radiation energy. What has not been tested, however, is the agreement of the actual emission footprints between ZHAireS and CoREAS. Differences are not excluded, as for example there are differences in the treatment of the electromagnetic cascade in the underlying CORSIKA and Aires codes \cite{2001ICRC....2..526H}. Direct comparisons of experimental data and ZHAireS simulations have so far been carried out only in a few selected cases \cite{HuegePLREP}.

At frequencies beyond 100~MHz, some measurements exist, e.g., by the LOFAR high-band antennas \cite{Nelles:2014dja}, by ARIANNA \cite{Barwick:2016mxm}, or by ANITA for which a determination of the cosmic-ray flux using ZHAireS simulations was in agreement with Auger data within uncertainties \cite{Schoorlemmer:2015ujs}. Still, the simulations have certainly been probed much less at high frequencies than at lower frequencies. This should be kept in mind and comparisons should be made once adequate data are available.

Finally, there have been measurements in the lab with the SLAC T-510 experiment, confirming that the pulse shapes and absolute amplitudes predicted with an implementation of the endpoints and ZHS formalisms in GEANT4 are in agreement with the lab measurement \cite{Bechtol:2021tyd}, again underlining the reliability of the microscopic simulation approach.


\section{Simulations for inclined air showers}

In the past few years, radio detection of inclined air showers, i.e., those with zenith angles beyond $\sim 60^\circ$ has become a particular focus of research. This is because these showers can be detected with sparse radio arrays and thus the radio technique can be applied up to the highest cosmic-ray energies \cite{HuegeUHECR2014,SchlueterARENA2022}.

The main difference between inclined and vertical air showers is that inclined showers develop to their maximum much higher in the atmosphere, so at much larger source distances from the antennas (dozens to over a hundred kilometres as opposed to a few kilometers), and thus also in a much less dense atmosphere. As a consequence, geometrical early-late effects start to become important and need to be considered \cite{Schluter:2022mhq}. Also, CoREAS simulations predict a refractive displacement of the radio emission with respect to the particle distributions \cite{Schluter:2020tdz}, which needs to be taken into account when analysing hybrid particle and radio data. Both of these effects have yet to be tested explicitly with experimental data.

\begin{figure}
\centering
\includegraphics[width=0.4\textwidth]{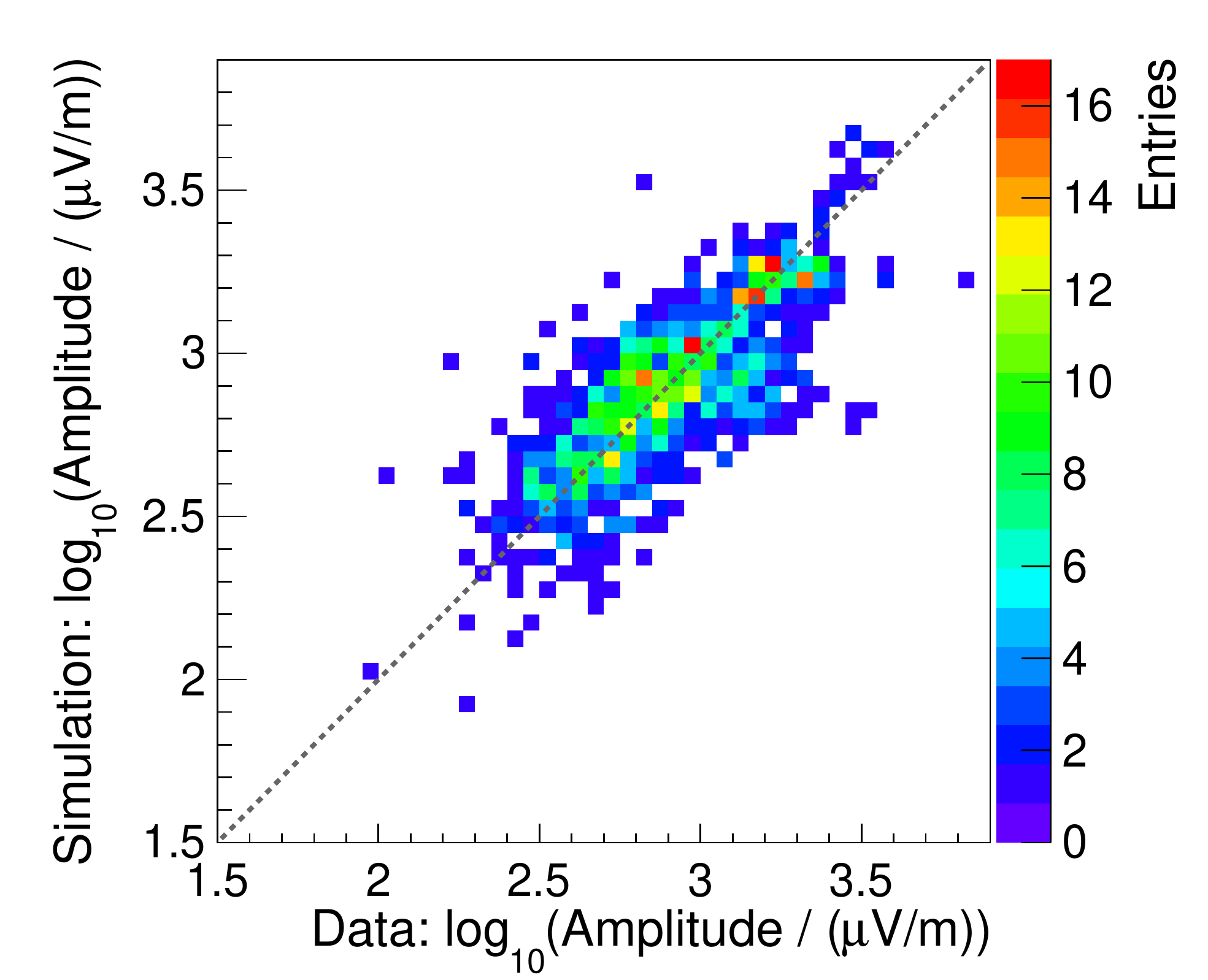}
\includegraphics[width=0.59\textwidth]{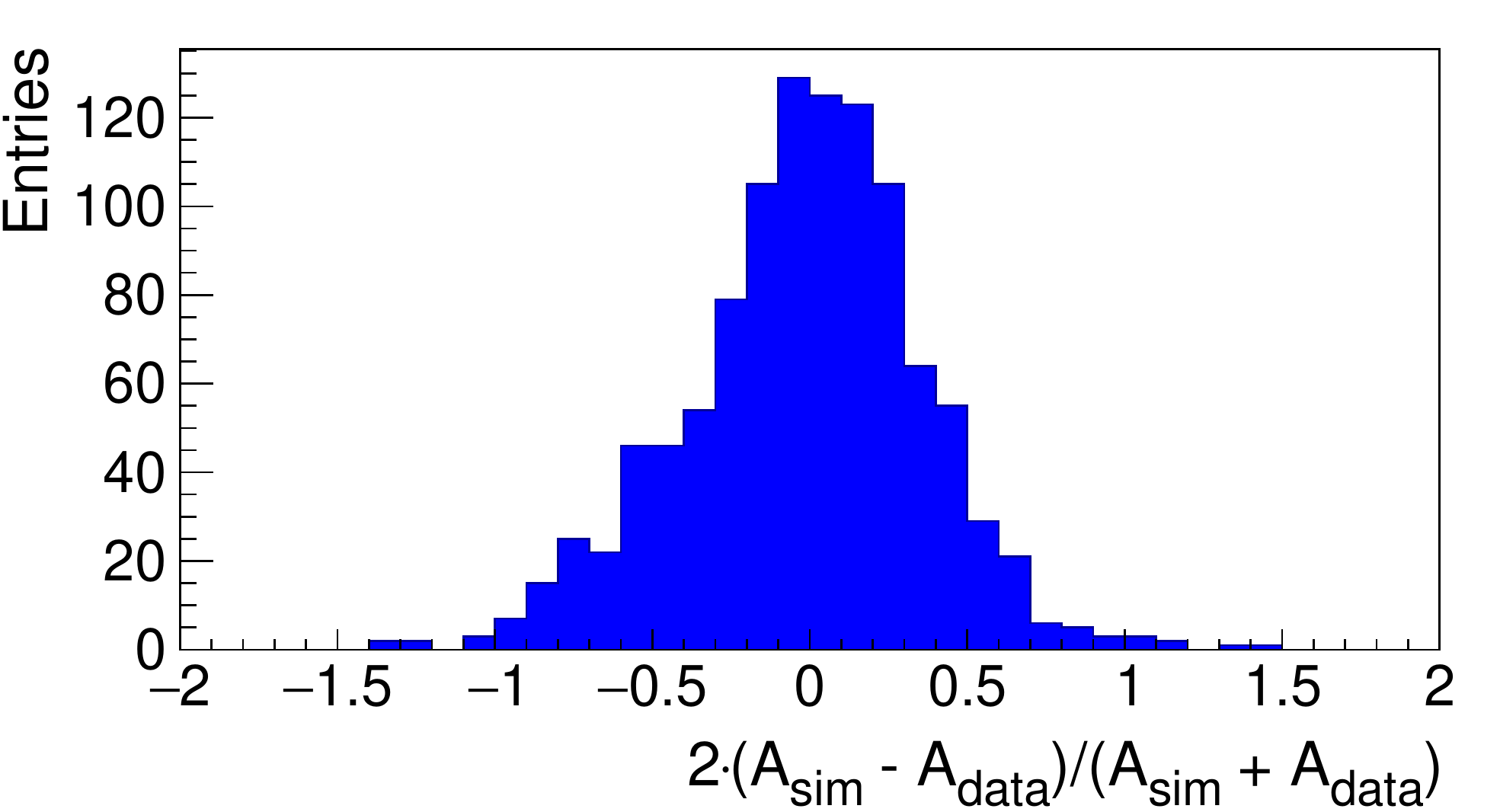}
\caption{\label{fig:aerahas}Agreement between CoREAS-simulated amplitudes for radio emission from inclined air showers measured by AERA. The scatter plot (left) illustrates a tight correlation between measurements and simulations. The difference histogram (right) shows agreement within 2\%, the joint scatter amounts to 38\%. From \cite{PierreAuger:2018pmw}.}
\end{figure}

The only experiment that has so far measured the radio emission of inclined air showers with relevant statistics is AERA, which has confirmed the simulation prediction that inclined air showers illuminate very large areas on the ground (due to the beamed emission from a far-away source) \cite{PierreAuger:2018pmw}. Furthermore, for a subset of 50 events it was shown that the CoREAS simulation predictions of the energy fluence in the 30--80~MHz band and the measurements with AERA were good agreement, see Fig.\ \ref{fig:aerahas}. At least up to zenith angles of $\sim 80^\circ$ and in the frequency-band from 30--80~MHz, we can thus have confidence in the validity of the simulations.

However, there are certainly still bigger uncertainties in the validation of simulations for inclined air showers than there are for vertical ones. When going to yet more inclined air showers, say beyond 85$^\circ$ zenith angles, near-surface propagation effects, which are currently completely ignored in existing simulations, could potentially play a role. In this context, it should also be stressed that current air-shower simulation codes assume that the radio emission propagates along straight lines in the atmosphere, whereas it is well known that the refractive index gradient curves the rays, which can lead to very interesting optical effects in inhomogeneous atmospheres \cite{SERON200595}. Work in that direction is currently underway \cite{DiederARENA2022}, with some focus on testing whether approximations made in current simulation codes may break down in the most extreme geometries.

\begin{figure}
\centering
\includegraphics[width=0.99\textwidth]{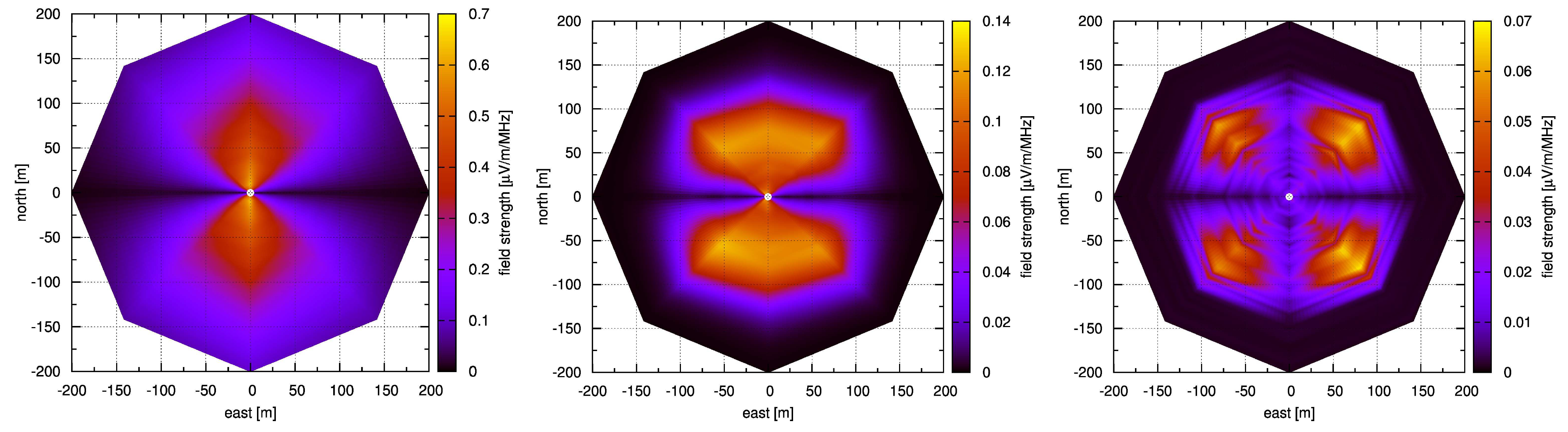}
\caption{\label{fig:coreascloverleafs}CoREAS-simuated radio-emission footprints showing the maximum electric field amplitudes in the $\vec{v} \times (\vec{v} \times \vec{B})$ polarisation for the 40--80~MHz (left), 300--1200~MHz (middle) and 3.4--4.2~GHz bands. This is for a vertical air shower with an energy of $10^{17}$\,eV induced by an iron primary at the site of the LOPES and CROME experiments (110 m above sea-level, geomagnetic field of 48 nT with 65$^\circ$ inclination). At high frequencies, a ``clover-leaf pattern'', indicating the presence of synchrotron emission, is apparent. From \cite{Huege:2013yra}.}
\end{figure}

Finally, recent simulation studies have revealed that for inclined air showers in magnetic fields stronger than the one at the site of the Pierre Auger Observatory, we can expect two important additional effects: a) a loss of coherence related to the fact that particle distributions become much wider in the less dense atmosphere at high altitudes \cite{ChicheARENA2022} and b) a transition from the ``transverse current'' regime to the ``synchrotron regime'' \cite{James:2022mea,ChicheARENA2022}. The presence of synchrotron emission in the radio signals had been predicted at GHz frequencies already early on \cite{Huege:2013yra}, manifesting itself as a ``clover-leaf pattern'' in the $\vec{v} \times (\vec{v} \times \vec{B})$ polarisation \cite{HuegeThesis2004}, and was possibly already seen by the CROME experiment \cite{CROMEPRL}. But for inclined air showers the synchrotron emission  seems to become relevant already at frequencies as low as 30--80~MHz. This means that our ``standard paradigm'' of ``transverse current plus charge-excess emission'' is likely to break down for very inclined air showers, in particular in magnetic fields stronger than those in the south-atlantic anomaly. While the Auger Radio detector might thus not be affected by these effects, for GRAND \cite{Alvarez-Muniz:2018bhp} sites in China this transition will have to be taken into account. To be clear, though, at this moment these are pure simulation predictions -- untested by experimental data.


\section{Future directions}

Microscopic Monte Carlo simulations with CoREAS and ZHAireS have served the community very well. However, we are also running into limitations of our current simulation approaches.

One problem is the ever-increasing amount of data we need to handle, with increasing event statistics and also increasing data quality. For example, LOFAR 2.0 is expected to increase LOFAR statistics by an order of magnitude. Scaling up current reconstruction procedures, which involve the simulation of many showers per individual measured event, will likely become prohibitive in the future. Also, the carbon footprint of running extensive simulations is a problem. SKA with more than 60,000 antennas in a dense core of 500~m diameter \cite{HuegeSKAIcrc2015,BuitinkARENA2022} will take this challenge to an even higher level.

Macroscopic approaches can possibly be used for specific purposes, but they have so far not been successful at correctly describing the emission with sufficient accuracy over the whole footprint. Here, ``hybrid approaches'' which use a full microscopic simulation of the air-shower radio emission as a basis but then scale it according to our understanding of the radio-emission physics could be a way out. For inclined air showers -- with a far-away source -- the ``radio morphing'' approach \cite{Chiche:2021iin} promises to make showers re-usable. For vertical air showers -- with a close source, incurring additional signal complexity -- the ``template synthesis'' approach \cite{DesmetARENA2022} has been shown to achieve an accuracy of $\sim 10$\%; it is currently undergoing generalization to other geometries than purely vertical.

Finally, the radio detection community has also been working hard on making in-ice radio detection of neutrino-induced particle showers a success. This also requires a proper modelling of the radio signals seen in the ice from both air showers and the cascades induced at the air-ice boundary. Currently, this is being studied by combining a modified version of CoREAS with a GEANT4-based simulation of the in-ice cascade \cite{LatifARENA2022,deKockereARENA2022}. In the future, CORSIKA~8 \cite{Huege:2022xbo} will be the natural framework in which such complex scenarios can be handled in one consistent simulation.


\section{Conclusions}

Microscopic simulations of radio emission from extensive air showers have served the community well. To date, no problems or inaccuracies have been found with the simulations. This is true in particular for vertical air showers, for which the simulations have been compared extensively with data, especially in the 30--80~MHz band. At higher frequencies, the simulations have been tested considerably less. Also, CoREAS has been compared with much more experimental data than ZHAireS, and while ZHAireS and CoREAS agree well on the prediction of the absolute radiation energy, the shapes of the emission footprints have not been compared in detail.

The understanding of the radio emission from inclined air showers has grown significantly in recent years, including geometrical early-late effects and an expected refractive displacement. However, only limited comparisons with experimental data could be made so far. While we have no indication of problems, a reliable validation is yet to be made. Furthermore, for very inclined air showers, say beyond $\sim 85^\circ$ zenith angle, near-surface propagation effects and/or straight-line approximations made in the simulation codes could turn out to raise problems. Finally, recent simulation studies predict a loss of coherence and the ``return of geosynchrotron emission'' for inclined air showers. These predictions need to be verified experimentally, and would require rethinking the paradigm adopted for decomposing the emission into geomagnetic and charge-excess radiation adopted so far.


\let\oldbibliography\thebibliography
\renewcommand{\thebibliography}[1]{%
  \oldbibliography{#1}%
  \setlength{\itemsep}{1pt}%
}

{\footnotesize

\providecommand{\href}[2]{#2}\begingroup\raggedright\endgroup
}
\end{document}

%% file: definitions.tex
\newcommand{\ev}[1]{\ensuremath{10^{#1}\,\text{eV}}\xspace}
\newcommand{\gcm}{\ensuremath{\text{g}\,\text{cm}^{-2}}\xspace}
\newcommand{\kgm}{\ensuremath{\text{kg}\,\text{m}^{-3}}\xspace}
\newcommand{\D}{\ensuremath{^\circ}\xspace}

\newcommand{\Eem}{\ensuremath{E_\mathrm{em}}\xspace}
\newcommand{\EemMC}{\ensuremath{E_\mathrm{em}^\mathrm{MC}}\xspace}
\newcommand{\Egeo}{\ensuremath{E_\mathrm{geo}}\xspace}
\newcommand{\Sgeo}{\ensuremath{S_\mathrm{geo}}\xspace}
\newcommand{\dmax}{\ensuremath{d_\mathrm{max}}\xspace}
\newcommand{\Xmax}{\ensuremath{X_\mathrm{max}}\xspace}
\newcommand{\xmax}{\ensuremath{X_\mathrm{max}}\xspace}
\newcommand{\rhomax}{\ensuremath{\rho_\mathrm{max}}\xspace}

\newcommand{\fgeo}{\ensuremath{f_\mathrm{geo}}\xspace}
\newcommand{\fce}{\ensuremath{f_\mathrm{ce}}\xspace}
\newcommand{\fgs}{\ensuremath{f_\mathrm{GS}}\xspace}
\newcommand{\ace}{\ensuremath{a_\mathrm{ce}}\xspace}

\newcommand{\vB}{\ensuremath{\vec{v} \times \vec{B}}\xspace}
\newcommand{\vxB}{\ensuremath{\vec{v} \times \vec{B}}\xspace}

\newcommand{\vvB}{\ensuremath{\vec{v} \times  (\vec{v} \times \vec{B})}\xspace}
\newcommand{\vxvxB}{\ensuremath{\vec{v} \times  (\vec{v} \times \vec{B})}\xspace}

\newcommand{\fvB}{\ensuremath{f_{\vec{v} \times \vec{B}}}\xspace}
\newcommand{\fvxB}{\ensuremath{f_{\vec{v} \times \vec{B}}}\xspace}
\newcommand{\fvvB}{\ensuremath{f_{\vec{v} \times  (\vec{v} \times \vec{B})}}\xspace}
\newcommand{\fv}{\ensuremath{f_{\vec{v}}}\xspace}

%% file: main.bbl
\begin{thebibliography}{10}

\bibitem{HuegePLREP}
T.~Huege, \emph{{Radio detection of cosmic ray air showers in the digital
  era}},
  \href{https://doi.org/http://dx.doi.org/10.1016/j.physrep.2016.02.001}{\emph{Physics
  Reports} {\bfseries 620} (2016) 1 }.

\bibitem{JamesFalckeHuege2012}
C.W.~James, H.~Falcke, T.~Huege and M.~Ludwig, \emph{{General description of
  electromagnetic radiation processes based on instantaneous charge
  acceleration in ``endpoints''}},
  \href{https://doi.org/10.1103/PhysRevE.84.056602}{\emph{Phys. Rev. E}
  {\bfseries 84} (2011) 056602}.

\bibitem{Huege:2013vt}
T.~Huege, M.~Ludwig and C.W.~James, \emph{{Simulating radio emission from air
  showers with CoREAS}}, \href{https://doi.org/10.1063/1.4807534}{\emph{AIP
  Conf. Proc.} {\bfseries 1535} (2013) 128}
  [\href{https://arxiv.org/abs/1301.2132}{{\ttfamily 1301.2132}}].

\bibitem{ZasHalzenStanev1992}
E.~{Zas}, F.~{Halzen} and T.~{Stanev}, \emph{{Electromagnetic pulses from
  high-energy showers: Implications for neutrino detection}}, {\emph{Phys. Rev.
  D} {\bfseries 45} (1992) 362}.

\bibitem{AlvarezMunizCarvalhoZas2012}
J.~{Alvarez-Mu\~niz}, W.R.~{Carvalho Jr.} and E.~{Zas}, \emph{{Monte Carlo
  simulations of radio pulses in atmospheric showers using ZHAireS}},
  \href{https://doi.org/10.1016/j.astropartphys.2011.10.005}{\emph{Astropart.
  Phys.} {\bfseries 35} (2012) 325 }.

\bibitem{LOFARNatureXmax}
S.~Buitink, A.~Corstanje, H.~Falcke and {et al.}, \emph{{A large light-mass
  component of cosmic rays at 10$^{17}$ -- 10$^{17.5}$ eV from radio
  observations}}, {\emph{Nature} {\bfseries 531} (2016) 70}.

\bibitem{Scholten:2016gmj}
O.~Scholten et~al., \emph{{Measurement of the circular polarization in radio
  emission from extensive air showers confirms emission mechanisms}},
  \href{https://doi.org/10.1103/PhysRevD.94.103010}{\emph{Phys. Rev. D}
  {\bfseries 94} (2016) 103010}
  [\href{https://arxiv.org/abs/1611.00758}{{\ttfamily 1611.00758}}].

\bibitem{Mitra:2020mza}
P.~Mitra et~al., \emph{{Reconstructing air shower parameters with LOFAR using
  event specific GDAS atmosphere}},
  \href{https://doi.org/10.1016/j.astropartphys.2020.102470}{\emph{Astropart.
  Phys.} {\bfseries 123} (2020) 102470}
  [\href{https://arxiv.org/abs/2006.02228}{{\ttfamily 2006.02228}}].

\bibitem{Tunka-Rex:2015zsa}
{\scshape Tunka-Rex} collaboration, \emph{{Radio measurements of the energy and
  the depth of the shower maximum of cosmic-ray air showers by Tunka-Rex}},
  \href{https://doi.org/10.1088/1475-7516/2016/01/052}{\emph{JCAP} {\bfseries
  01} (2016) 052} [\href{https://arxiv.org/abs/1509.05652}{{\ttfamily
  1509.05652}}].

\bibitem{AERAEnergyPRL}
{\scshape Pierre Auger Collaboration} collaboration, \emph{{Measurement of the
  Radiation Energy in the Radio Signal of Extensive Air Showers as a Universal
  Estimator of Cosmic-Ray Energy}},
  \href{https://doi.org/10.1103/PhysRevLett.116.241101}{\emph{Phys. Rev. Lett.}
  {\bfseries 116} (2016) 241101}.

\bibitem{PontXmaxARENA2022}
{\scshape B. Pont for the Pierre Auger} collaboration, \emph{{The depth of the
  shower maximum of air showers measured with AERA}}, {\emph{these proceedings}
  (2023) }.

\bibitem{Gottowik:2017wio}
M.~Gottowik, C.~Glaser, T.~Huege and J.~Rautenberg, \emph{{Determination of the
  absolute energy scale of extensive air showers via radio emission: systematic
  uncertainty of underlying first-principle calculations}},
  \href{https://doi.org/10.1016/j.astropartphys.2018.07.004}{\emph{Astropart.
  Phys.} {\bfseries 103} (2018) 87}
  [\href{https://arxiv.org/abs/1712.07442}{{\ttfamily 1712.07442}}].

\bibitem{2001ICRC....2..526H}
D.~{Heck}, J.~{Knapp} and S.J.~{Sciutto}, \emph{{Study of model dependence of
  EAS simulations at E~$\geq10^{19}$~eV.}},  in \emph{Proceedings of the 27th
  International Cosmic Ray Conference, Hamburg, Germany}, vol.~2, p.~526, 2002.

\bibitem{Nelles:2014dja}
A.~Nelles et~al., \emph{{Measuring a Cherenkov ring in the radio emission from
  air showers at 110\textendash{}190 MHz with LOFAR}},
  \href{https://doi.org/10.1016/j.astropartphys.2014.11.006}{\emph{Astropart.
  Phys.} {\bfseries 65} (2015) 11}
  [\href{https://arxiv.org/abs/1411.6865}{{\ttfamily 1411.6865}}].

\bibitem{Barwick:2016mxm}
S.W.~Barwick et~al., \emph{{Radio detection of air showers with the ARIANNA
  experiment on the Ross Ice Shelf}},
  \href{https://doi.org/10.1016/j.astropartphys.2017.02.003}{\emph{Astropart.
  Phys.} {\bfseries 90} (2017) 50}
  [\href{https://arxiv.org/abs/1612.04473}{{\ttfamily 1612.04473}}].

\bibitem{Schoorlemmer:2015ujs}
H.~Schoorlemmer et~al., \emph{{Energy and Flux Measurements of Ultra-High
  Energy Cosmic Rays Observed During the First ANITA Flight}},
  \href{https://doi.org/10.22323/1.236.0272}{\emph{PoS} {\bfseries ICRC2015}
  (2016) 272}.

\bibitem{Bechtol:2021tyd}
K.~Bechtol et~al., \emph{{SLAC T-510 experiment for radio emission from
  particle showers: Detailed simulation study and interpretation}},
  \href{https://doi.org/10.1103/PhysRevD.105.063025}{\emph{Phys. Rev. D}
  {\bfseries 105} (2022) 063025}
  [\href{https://arxiv.org/abs/2111.04334}{{\ttfamily 2111.04334}}].

\bibitem{HuegeUHECR2014}
T.~{Huege} and A.~{Haungs}, \emph{{Radio detection of cosmic rays: present and
  future}}, {\emph{JPS Conference Proceedings} {\bfseries 09} (2016) 010018}.

\bibitem{SchlueterARENA2022}
{\scshape F. Schlueter for the Pierre Auger} collaboration, \emph{{Expected
  performance of the Auger Radio Detector}}, {\emph{these proceedings} (2023)
  }.

\bibitem{Schluter:2022mhq}
F.~Schl\"uter and T.~Huege, \emph{{Signal model and event reconstruction for
  the radio detection of inclined air showers}}, {\emph{JCAP in press} (2022) }
  [\href{https://arxiv.org/abs/2203.04364}{{\ttfamily 2203.04364}}].

\bibitem{Schluter:2020tdz}
F.~Schl\"uter, M.~Gottowik, T.~Huege and J.~Rautenberg, \emph{{Refractive
  displacement of the radio-emission footprint of inclined air showers
  simulated with CoREAS}},
  \href{https://doi.org/10.1140/epjc/s10052-020-8216-z}{\emph{Eur. Phys. J. C}
  {\bfseries 80} (2020) 643}
  [\href{https://arxiv.org/abs/2005.06775}{{\ttfamily 2005.06775}}].

\bibitem{PierreAuger:2018pmw}
{\scshape Pierre Auger} collaboration, \emph{{Observation of inclined EeV air
  showers with the radio detector of the Pierre Auger Observatory}},
  \href{https://doi.org/10.1088/1475-7516/2018/10/026}{\emph{JCAP} {\bfseries
  10} (2018) 026} [\href{https://arxiv.org/abs/1806.05386}{{\ttfamily
  1806.05386}}].

\bibitem{SERON200595}
F.~Seron, D.~Gutierrez, G.~Gutierrez and E.~Cerezo, \emph{Implementation of a
  method of curved ray tracing for inhomogeneous atmospheres},
  \href{https://doi.org/https://doi.org/10.1016/j.cag.2004.11.010}{\emph{Computers
  \& Graphics} {\bfseries 29} (2005) 95}.

\bibitem{DiederARENA2022}
D.~van~den Broeck et~al., \emph{{Radio propagation in non-uniform media}},
  {\emph{these proceedings} (2023) }.

\bibitem{Huege:2013yra}
T.~Huege and C.W.~James, \emph{{Full Monte Carlo simulations of radio emission
  from extensive air showers with CoREAS}},  in \emph{{33rd International
  Cosmic Ray Conference}}, p.~0548, 7, 2013
  [\href{https://arxiv.org/abs/1307.7566}{{\ttfamily 1307.7566}}].

\bibitem{ChicheARENA2022}
S.~Chiche et~al., \emph{{New features in the radio-emission of very inclined
  showers}}, {\emph{these proceedings} (2023) }.

\bibitem{James:2022mea}
C.W.~James, \emph{{Nature of radio-wave radiation from particle cascades}},
  \href{https://doi.org/10.1103/PhysRevD.105.023014}{\emph{Phys. Rev. D}
  {\bfseries 105} (2022) 023014}
  [\href{https://arxiv.org/abs/2201.01298}{{\ttfamily 2201.01298}}].

\bibitem{HuegeThesis2004}
T.~{Huege}, \emph{{Radio Emission from Cosmic Ray Air Showers}}, Ph.D. thesis,
  Rheinische Friedrich-Wilhelms-Universit\"at Bonn, Germany, 2004.

\bibitem{CROMEPRL}
R.~{\v{S}m\'{i}da}, F.~{Werner}, R.~{Engel} and {et al.}, \emph{{First
  Experimental Characterization of Microwave Emission from Cosmic Ray Air
  Showers}}, \href{https://doi.org/10.1103/PhysRevLett.113.221101}{\emph{Phys.
  Rev. Lett.} {\bfseries 113} (2014) 221101}.

\bibitem{Alvarez-Muniz:2018bhp}
{\scshape GRAND} collaboration, \emph{{The Giant Radio Array for Neutrino
  Detection (GRAND): Science and Design}},
  \href{https://doi.org/10.1007/s11433-018-9385-7}{\emph{Sci. China Phys. Mech.
  Astron.} {\bfseries 63} (2020) 219501}
  [\href{https://arxiv.org/abs/1810.09994}{{\ttfamily 1810.09994}}].

\bibitem{HuegeSKAIcrc2015}
T.~{Huege}, J.~{Bray}, S.~{Buitink} and {et al.}, \emph{{High-precision
  measurements of extensive air showers with the SKA}},  in \emph{Proceedings
  of the 34th International Cosmic Ray Conference, The Hague, The Netherlands},
  PoS(ICRC2015)309.

\bibitem{BuitinkARENA2022}
S.~Buitink et~al., \emph{{Constraining the cosmic-ray mass composition by
  measuring the shower length with SKA}}, {\emph{these proceedings} (2023) }.

\bibitem{Chiche:2021iin}
S.~Chiche, O.~Martineau-Huynh, K.~Kotera, M.~Tueros and K.~D.~de Vries,
  \emph{{Radio-Morphing: a fast, efficient and accurate tool to compute the
  radio signals from air-showers}},
  \href{https://doi.org/10.22323/1.395.0194}{\emph{PoS} {\bfseries ICRC2021}
  (2021) 194} [\href{https://arxiv.org/abs/2202.05886}{{\ttfamily
  2202.05886}}].

\bibitem{DesmetARENA2022}
M.~Desmet et~al., \emph{{Template synthesis approach for radio emission from
  extensive air showers}}, {\emph{these proceedings} (2023) }.

\bibitem{LatifARENA2022}
U.~Latif et~al., \emph{{Propagating Air Shower Radio Signals to In-ice
  Antennas}}, {\emph{these proceedings} (2023) }.

\bibitem{deKockereARENA2022}
S.~de~Kockere et~al., \emph{{Simulation of the propagation of cosmic ray air
  shower cores in ice}}, {\emph{these proceedings} (2023) }.

\bibitem{Huege:2022xbo}
{\scshape CORSIKA} collaboration, \emph{{CORSIKA 8 -- the next-generation air
  shower simulation framework}},  in \emph{{21st International Symposium on
  Very High Energy Cosmic Ray Interactions}}, 8, 2022
  [\href{https://arxiv.org/abs/2208.14240}{{\ttfamily 2208.14240}}].

\end{thebibliography}
